\documentclass[fdp,fleqn]{w-art}
\usepackage{times}
\usepackage{w-thm}
\usepackage{amsmath}
\usepackage{amscd}
\usepackage{amsfonts}
\usepackage[]{graphicx}
\newcommand{\ee}{\end{eqnarray}}
\newcommand{\be}{\begin{eqnarray}}

\newcommand{\coker}{\operatorname{coker}}
\newcommand{\rref}[1]{(\ref{#1})}

\begin{document}
\DOIsuffix{theDOIsuffix}
\Volume{55}
\Issue{1}
\Month{01}
\Year{2007}
\pagespan{1}{}



\title[D-branes on generalized K\"ahler geometries: I. General formalism]{N = 2 world-sheet approach to D-branes on generalized K\"ahler geometries: I. General formalism}


\author[A. Sevrin]{Alexander Sevrin\inst{1,}%
  \footnote{E-mail: Alexandre.Sevrin@vub.ac.be}}
\address[\inst{1}]{Theoretische Natuurkunde, Vrije Universiteit Brussel and
The International Solvay Institutes,\\
Pleinlaan 2, B-1050 Brussels, Belgium.}
\author[W. Staessens]{Wieland Staessens\inst{1,}\footnote {Aspirant FWO, E-mail: Wieland.Staessens@vub.ac.be}}
\author[A. Wijns]{Alexander Wijns\inst{2,3,}\footnote{E-mail: awijns@nordita.org}}
\address[\inst{2}]{Department of Mathematics,
Science Institute, University of Iceland,\\
Dunhaga 2, 107 Reykjavik, Iceland.}
\address[\inst{3}]{NORDITA,
Roslagstullsbacken 23, 106 91 Stockholm, Sweden.}
\begin{abstract}
We present an $N=2$ world-sheet superspace description of D-branes on bihermitian or generalized K\"ahler manifolds. To accomplish this, D-branes are considered as boundary conditions for a nonlinear $\sigma$-model in what we call $N=2$ boundary superspace. In this note the general formalism for such an approach is presented and the resulting classification sketched. This includes some remarks regarding target spaces whose parameterization includes semi-chiral superfields which have not appeared in the literature yet. In an accompanying note we turn to some examples and applications of the general setup presented here. 
\end{abstract}
\maketitle                   





\section{Introduction}

It has long been known that there is a deep and rich connection between supersymmetric nonlinear $\sigma$-models and complex geometry. From a string theory point of view, a two dimensional nonlinear $\sigma$-model -- with as basic objects maps $X: \Sigma \rightarrow {\cal M}$ -- corresponds to the field theory on the world-sheet $\Sigma$ of a string, which is embedded into some target space ${\cal M}$ by an embedding map $X$. It is the existence of extended supersymmetry on $\Sigma$ which then relates to additional structure on ${\cal M}$. Here we focus on $N=(2,2)$ supersymmetry. Apart from the obvious motivation coming from string theory, this is clearly a problem worthy of study in its own right. Especially so when, again motivated by string theory, the target space is not simply a Riemannian manifold endowed with a metric $g$, but is also equipped with a torsionful connection, where the torsion is derived from an NS-NS three-form $H$.  Usually, a superspace description of the world-sheet theory helps unraveling this intricate connection with geometrical structures on ${\cal M}$

In $N=(2,2)$ superspace on a surface without boundary, corresponding to the world-sheet of a closed string, the situation is by now quite well understood: the target space geometry allowing for $N=(2,2)$ world-sheet supersymmetry is bihermitian \cite{Gates:1984nk} or -- in the language of generalized complex geometry -- generalized K\"ahler \cite{Gualtieri:2003dx}, and can always be parameterized by at least one of three types of $N=(2,2)$ superfield called chiral, twisted chiral and semi-chiral superfields \cite{Lindstrom:2005zr}.

Things are still much less understood for world-sheets with boundaries, corresponding to open strings, where the boundary of $\Sigma$ is mapped to a submanifold ${\cal N}$ of ${\cal M}$ wrapped by a D-brane. The presence of a boundary breaks at least part of the world-sheet supersymmetry, and it is the case where half of the supersymmetry is preserved that is important for understanding D-branes which are BPS in target space. We are thus led to the study of open string boundary conditions in a bihermitian background which preserve half of the world-sheet supersymmetry. It is however important to note that solving for these boundary conditions does not necessarily lead to BPS D-branes. To achieve this, the boundary conditions should also respect quantum conformal invariance. Exactly this fact also serves as an important motivation for understanding the extended superspace description of D-branes. Indeed, one way of acquiring quantum conformal invariance is by imposing the world-sheet $\beta$-functions to vanish. Achieving this at higher order in perturbation theory involves higher loop quantum field theory calculations, and is greatly facilitated by having a superspace formulation at hand \cite{Grisaru:1986px,Nevens:2006ht}.

Although superspace approaches to $\sigma$-models with boundaries have been the subject of much investigation, an appropriate superspace formulation is still lacking. Here, we review how the notion of a boundary superspace promises to resolve this problem. $N=1$ boundary superspace was introduced and explored in \cite{Koerber:2003ef}. In \cite{Sevrin:2007yn} an $N=2$ boundary superspace formalism was used to recover the full classification of A and B branes on K\"ahler manifolds, which are parameterized by either chiral or twisted chiral superfields exclusively. Especially for coisotropic A branes \cite{Kapustin:2001ij,Lindstrom:2002jb} the solution proved to be quite subtle and elegant. In \cite{Sevrin:2008tp} this was taken one step further to cover bihermitian geometries with commuting complex structures, which generically have a local description in terms of chiral and twisted chiral fields simultaneously. In this note we also discuss some interesting aspects of the classification when semi-chiral superfields are included. The full classification will appear elsewhere \cite{wip}. 

For a far more detailed account of these matters and a more complete list of references, we invite the reader to consult \cite{Sevrin:2007yn,Sevrin:2008tp}. Here we will only provide a rough sketch of the boundary superspace formalism and the way it is used to classify $N=2$ boundary conditions. But first we briefly summarize some facts about the relation between supersymmetry and geometry in the absence of world-sheet boundaries.

\section{$N=(2,2)$ superspace and geometry}

Given a target space ${\cal M}$ with metric $g$ and three-form $H$, the dynamics of a closed supersymmetric string propagating in such a background is encoded in the action
 \be
 {\cal S}=2\int d^2 \sigma \, d \theta^+ d\theta^- \,D_+X^aD_-X^b\left(g_{ab}+b_{ab} \right),\label{an11}
 \ee
where we used that locally $H = db$. Here we wrote down the action in $N=(1,1)$ superspace, which required the introduction of two real Grassmann coordinates $\theta^\pm$ on top of the usual (bosonic) world-sheet coordinates $\tau$ and $\sigma$. The corresponding supercovariant derivatives are denoted by $D_\pm$ and satisfy a standard superalgebra; see for example \cite{Sevrin:2008tp}. Furthermore $X^a$, $g$ and $b$ are $N=(1,1)$ superfields. 

The $N=(1,1)$ superspace action eq. \rref{an11} can be written down for any Riemannian manifold $({\cal M}, g)$. One can however ask for what kind of backgrounds this action exhibits additional supersymmetries. For $N=(2,2)$ supersymmetry, this implies that there exist additional symmetries of the form
\begin{eqnarray}
\delta X^a= \varepsilon ^+\,J_+^a{}_b(X)\,D_+X^b+\varepsilon ^-\,J_-^a{}_b(X)\,D_-X^b,\label{tr22}
\end{eqnarray}
which are of the most general form consistent with dimensions and super Poincar\'e invariance. Here, $\varepsilon^\pm$ are real Grassmann supersymmetry parameters and $J_\pm$ are a priori arbitrary (1,1)-tensors which are allowed to depend on the superfields $X^a$. On-shell closure of the $N=(2,2)$ algebra and invariance of \rref{an11} under \rref{tr22} however require that $J_\pm$ are complex structures, that $g$ is hermitian with respect to both of them and that they are covariantly constant with respect to a torsionful connection: 
$\nabla_\pm J_\pm = 0$, where $\nabla_\pm = \nabla_g \pm g^{-1}H$,
and $\nabla_g$ is the Levi-Civita connection. Note that hermiticity of the metric with respect to $J_\pm$ implies the existence of two two-forms $\omega_\pm = g J_\pm$, which are however not closed when $H$ is nonzero. Hence, these geometries are K\"ahler only when $H=0$. More generally, such geometries are called bihermitian.

The result of the previous paragraph has been known for over twenty years \cite{Gates:1984nk}, but only recently, a better understanding of such geometries has been achieved. The first reason for this was the development of generalized complex geometry (GCG) \cite{Hitchin:2004ut,Gualtieri:2003dx}. Indeed, in \cite{Gualtieri:2003dx} it was shown that bihermitian geometry is equivalent to what in the language of GCG is called generalized K\"ahler geometry. A generalized complex structure (GCS) is an automorphism ${\cal J}$ of $T{\cal M} \oplus T^\ast{\cal M}$ that squares to minus the identity, preserves the natural pairing defined on $T{\cal M} \oplus T^\ast{\cal M}$ and is involutive with respect to the Courant bracket. A generalized K\"ahler structure then requires the existence of two commuting GCSs ${\cal J}_1$ and ${\cal J}_2$, such that their product defines a definite inner product on $T{\cal M} \oplus T^\ast{\cal M}$. Up to a b-transform, the GCSs ${\cal J}_1$ and ${\cal J}_2$ are completely specified by the complex structures $J_\pm$, and $g$ and $H$,\footnote{Another approach is to twist the GCSs with respect to $H$ and put $b=0$ in their concrete expression given here.}
 \be
 {\cal J}_{1,2} = \frac 1 2
                                         \left( \begin{array}{cc}
                                        1 & 0 \\
                                        b & 1 
                                        \end{array}\right)
                                        \left( \begin{array}{cc}
                                        J_+ \pm J_- & -(\omega^{-1}_+ \mp \omega^{-1}_-) \\
                                        \omega_+ \mp \omega_- & -(J^t_+ \pm J^t_-) 
                                        \end{array}\right)
                                        \left( \begin{array}{cc}
                                        1 & 0 \\
                                        -b & 1 
                                        \end{array}\right).\label{gcs}
 \ee                         
It is easy to see that for instance for $J_+ = J_- = J$, the GCS ${\cal J}_1$ corresponds to a complex structure $J$ and ${\cal J}_2$ corresponds to a symplectic structure $w = gJ$. This example obviously corresponds to a K\"ahler structure. 

A second, related development was the establishment that the most general bihermitian target space can be parameterized as follows \cite{Sevrin:1996jr,Lindstrom:2005zr}. Locally, the target space can be decomposed as $\ker(J_+ - J_-) \oplus \ker(J_+ + J_-) \oplus \coker[J_+, J_-]$. From an $N=(2,2)$ superspace point of view, it turns out that each component of this decomposition can be parameterized by a different kind of superfield. To understand this better, let us introduce two more Grassmann coordinates $\hat\theta^\pm$ (and corresponding supercovariant derivatives $\hat D_\pm$) on top of $\theta^\pm$ to form an $N=(2,2)$ superspace. On dimensional grounds, the most general $N=(2,2)$ superspace action must necessarily be of the following form:
 \be
 {\cal S}=\int\,d^2 \sigma \,d\theta^+ \,d\theta^- \,d \hat \theta^+ \,d \hat \theta^- \, V(X), \label{an22}
 \ee
where $V$ is a real dimensionless scalar potential. To obtain some dynamics, one must impose constraint equations on the superfields. Imposing linear constraints of the form $\hat D_\pm X^a = J^a_\pm{}_b D X^b$, implies that $J_\pm$ are commuting complex structures. Thus, one expects such constrained superfields to parameterize $\ker[J_+, J_-]$. Indeed, diagonalizing $J_+$ and $J_-$ simultaneously and introducing complex coordinates with respect to $J_+$, we find that $\ker(J_+ - J_-)$ is parameterized by chiral fields $z^\alpha$ and anti-chiral fields $z^{\bar\alpha}$,
 \be
 \hat D_\pm z^\alpha = i D_\pm z^\alpha, \quad \hat D_\pm z^{\bar\alpha} = -i D_\pm z^{\bar\alpha},
 \ee
while $\ker(J_+ + J_-)$ is parameterized by twisted chiral fields $w^\mu$ and twisted anti-chiral fields $w^{\bar\mu}$,
 \be
 \hat D_\pm w^\mu = \pm i D_\pm w^\mu, \quad \hat D_\pm w^{\bar\mu} = \mp i D_\pm w^{\bar\mu}. \label{n22tc}
 \ee
These superfields have exactly the same number of components as $N=(1,1)$ superfields. On the other hand, since the supersymmetry algebra only closes off-shell up to terms involving $[J_+,J_-]$, we expect to need auxiliary fields in the $N=(2,2)$ parameterization of $\coker[J_+, J_-]$, and thus ``less constrained'' superfields in these directions. These superfields form a semi-chiral multiplet $l^{\tilde\alpha}$, $r^{\tilde\mu}$, and a semi-anti-chiral multiplet $l^{\bar{\tilde\alpha}}$ and $r^{\bar{\tilde\mu}}$, which obey
 \be
 \hat D_+ l^{\tilde\alpha} = i D_+ l^{\tilde\alpha}, &&\quad \hat D_+ l^{\bar{\tilde\alpha}} = -i D_+ l^{\bar{\tilde\alpha}},\\
  \hat D_- r^{\tilde\mu} = i D_- r^{\tilde\mu}, &&\quad \hat D_- r^{\bar{\tilde\mu}} = -i D_- r^{\bar{\tilde\mu}},
 \ee
where consistency requires an equal number of $l^{\tilde\alpha}$ and $r^{\tilde\mu}$. The result of \cite{Lindstrom:2005zr} was essentially that only the above types of superfield are required to parameterize the most general bihermitian target space. All necessary data defining the background, i.e. the complex structures $J_\pm$, the metric $g$ and the three-form $H$, can be derived from the single Lagrangian density $V$, which is consequently called the generalized K\"ahler potential. In the chiral and twisted chiral sectors, these data derive linearly from $V$, while in the semi-chiral sector, the analogous relations are highly nonlinear.

\section{Boundary superspace}

It has been argued that supersymmetric D-branes in generalized K\"ahler backgrounds are generalized complex submanifolds $({\cal N}, {\cal F} )$ -- where ${\cal N}$ is a submanifold and $d{\cal F} = H$ on ${\cal N}$ -- with respect to either ${\cal J}_1$ or ${\cal J}_2$ \cite{Gualtieri:2003dx,Zabzine:2004dp}. For K\"ahler manifolds parameterized by chiral fields, it is not difficult to see that with the conventions from the previous section, this notion with respect to ${\cal J}_1$ leads to B branes, while the same notion with respect to ${\cal J}_2$ leads to A branes  \cite{Gualtieri:2003dx}. Equivalently, one can consider only ${\cal J}_1$ and instead exchange chiral fields for twisted chiral fields to obtain A branes. It is this latter approach we will adopt in the boundary superspace formalism. Unfortunately, the geometric understanding of D-branes as generalized complex submanifolds is much less developed once $H$ differs from zero.  Nevertheless, using boundary superspace, we can give a fairly concrete local description of D-branes in bihermitian backgrounds. 

In the absence of boundaries, the action \rref{an11} is manifestly $N=(1,1)$ supersymmetric. However, in the presence of a boundary -- at $\sigma = 0$ say -- part of the translation invariance and therefore half of the world-sheet supersymmetry is necessarily broken. Introducing a new basis,
$D = D_+ + D_-$ and $D' = D_+ - D_-$,
we can choose the broken supersymmetry to correspond to $D'$, while $D$ is preserved.
It now turns out that the action \cite{Lindstrom:2002mc,Koerber:2003ef}
 \be
 {\cal S}=-\int d^2 \sigma \, d \theta
\,D'\left(D_+X^aD_-X^b\left(g_{ab}+b_{ab}\right)\right),\label{an1}
 \ee
differs from (\ref{an11}) only by a boundary term, while being manifestly invariant under the $N=1$ supersymmetry corresponding to $D$. This is thus called an action in $N=1$ boundary superspace. The boundary term in the variation of (\ref{an1}) disappears by imposing either Dirichlet boundary conditions, $\delta X^a = 0$, or Neumann boundary conditions, $D'X^a = b^a{}_b DX^b$. To describe branes of intermediate dimensions one introduces an almost product structure; see \cite{Sevrin:2008tp} and references therein. Notice that the action \rref{an1} is not unique, but one can add a boundary term of the form,
\be
{\cal S}_b=2i\,\int d \tau \, d \theta\, (A_a\,DX^a + B_a D' X^a). \label{Eva}
\ee
The first term simply leads to a replacement $b \rightarrow {\cal F} = b + F$, where $F=dA$, in \rref{an1}. The second term seems to be problematic at first sight. A term like this arises naturally when considering twisted chiral or semi-chiral superfields. It turns out that it however always reduces to the form of the first term in \rref{Eva} when appropriate boundary conditions (e.g. Neumann conditions $D'X = {\cal F} DX$) are imposed.

To similarly go from $N=(2,2)$ to $N=2$ superspace, we again introduce operators $D$, $D'$, $\hat D$ and $\hat D'$, where again
$\hat D = \hat D_+ + \hat D_-$ and $\hat D' = \hat D_+ - \hat D_-$.
We take $D$ and $\hat D$ to correspond to preserved supersymmetries, while the other combinations are broken. By the same construction that led us to the $N=1$ action, we find that the action
 \be
 {\cal S}=\int d^2 \sigma\, d \theta d \hat \theta\, D' \hat D'\, V(X, \bar X)+
 i\,\int d \tau \,d \theta d \hat \theta \,W( X, \bar X), \label{an2}
 \ee
has manifest $N=2$ supersymmetry and differs from (\ref{an22}) only by a boundary term. The symbol $X$ here stands collectively for chiral, twisted chiral and semi-chiral superfields. Note that we were able to add a term with a boundary potential $W$, which turns out to be crucial for the consistency of the formalism. For instance, eq. (\ref{an22}) is invariant under generalized K\"ahler transformations
 \be
 V \rightarrow V + F + \bar F + G + \bar G\,, \quad F \equiv F(z,w,l)\,, ~G \equiv G(z, \bar w, r).
 \ee
This invariance only survives in the presence of a boundary if at the same time $W$ transforms as
 \be
 W \rightarrow W -i( F - \bar F) +i (G - \bar G).
 \ee

Again, the variation of the action with respect to the superfields will result in a boundary term. $N=(2,2)$ supersymmetry puts strong conditions on which kind of Dirichlet and Neumann conditions can be imposed. This then results in a classification of the possible D-branes in a bihermitian background. To get a feeling for what kind of boundary conditions we can impose, we can already learn a lot by analyzing the $N=2$ superfield content one obtains starting from the constrained superfields in $N=(2,2)$ superspace. In this respect chiral fields turn out to be quite different from the other types of superfield, while twisted and semi-chiral superfields are in some important ways very similar. 

Starting from an $N=(2,2)$ chiral superfield $z$ and its complex conjugate $\bar z$, we end up with the $N=2$ superfields $z$, $\bar z$, $D'z$ and $D'\bar z$. Note however that in $N=2$ superspace these fields are still constrained. The situation is very different for a twisted chiral field $w$. From eq. \rref{n22tc} we get $N=2$ superspace relations like $\hat D w = i D' w$ which relate the $N=2$ superfields $w$ and $D'w$. This implies that in $N=2$ superspace, we only end up with the superfields $w$ and $\bar w$, which are unconstrained. Roughly the same happens for a semi-chiral multiplet $l$, $r$, $\bar l$ and $\bar r$: in $N=2$ superspace we end up with the unconstrained superfields $l$, $r$, $\bar l$ and $\bar r$, and some (unconstrained) auxiliary fields. 

Knowing this, a lot can already be understood about possible boundary conditions. Since for chiral fields $Dz$ and $D'z$ are unrelated, Dirichlet and Neumann conditions can be imposed independently, while every condition implies its complex conjugate. So in chiral directions, one finds an even number of Dirichlet and an even of Neumann boundary conditions. For example, on K\"ahler manifolds parameterized by chiral fields exclusively, one finds holomorphically embedded submanifolds, $[J,R] = 0$, with a holomorphic $U(1)$ bundle with field strength $F_{\alpha\bar\beta} = -i \partial_\alpha \partial_{\bar\beta} W$, namely the well known B branes. 

In contrast, for twisted chiral and semi-chiral fields, one has two options. First of all, imposing a Dirichlet condition on a twisted chiral field automatically implies a Neumann condition, basically because of the constraint equation $\hat D w = i D' w$. For semi-chiral fields it turns out that Dirichlet conditions always come in even numbers, while they still automatically imply (an even number) of Neumann conditions.\footnote{Note that this implies that D0-branes preserving $N=2$ world-sheet supersymmetry can only exist on K\"ahler manifolds, while D1-branes only exist on manifolds parameterized by one twisted chiral field, any number of chiral fields and their complex conjugates.} 
Restricting to the case where chiral fields are absent allows us to be a bit more concrete. The boundary term in the variation of the action can in that case be written as 
 \be
 \delta {\cal S}\vert_{bdy} = \int d\tau d\theta d\hat\theta \, (B_a \delta X^a + \delta W), \label{bv}
 \ee
where the $B_a$ are certain expressions involving the bulk potential $V$. The exterior derivative of the one-form with components $B_a$ turns out to be $\Omega$, the symplectic form which exists in the absence of chiral fields,
 \be
 \Omega = 4g (J_+ - J_-)^{-1}.
 \ee 
Let us take the extremal case where we impose a maximal number of Dirichlet conditions, so that ${\cal N}$ has half the dimensionality of ${\cal M}$. Introducing local coordinates $\sigma^i$ on the brane, the vanishing of \rref{bv} implies the condition,
$\partial_i W = - B_a \partial_i X^a$.
The integrability condition for this is precisely that the pull-back of the symplectic form $\Omega$ to the brane vanishes, so that the brane wraps a lagrangian cycle with respect to $\Omega$. 

On the other hand, since twisted and semi-chiral superfields are unconstrained at the boundary, we can choose to impose a boundary constraint on them. Taking again the extreme case where such a constraint is imposed on all $X^a$ in eq. \rref{bv}, we get
 \be
 \hat D X^a = K^a{}_b (X) D X^b, \label{bc}
 \ee
so that all $X^a$ become chiral at the boundary. Note that this results in a space-filling brane. As before, consistency of an equation like this immediately results in the fact that $K$ is a complex structure. One can show that $\Omega$ is a (2,0) + (0,2) form with respect to $K$, which implies that the target space needs to be $4n$-dimensional, where $n\in {\mathbb N}$. This type of brane also requires a non-vanishing world-volume flux of the form $F = \Omega K$. Summarizing, these are the conditions for a maximally coisotropic brane, again with respect to $\Omega$. The intermediate cases describe coisotropic branes of intermediate dimension. In the special case where only twisted chiral fields are present, this story reduces to the known one for A branes on K\"ahler manifolds \cite{Kapustin:2001ij}. Our results illustrate in a very concrete way  that for bihermitian geometries of symplectic type (although no longer necessarily K\"ahler), there exists a class of D-branes, which are generalized complex submanifolds with respect to the symplectic structure,\footnote{That ${\cal J}_1$ is indeed of symplectic type follows from the fact that $\omega_+^{-1} - \omega_-^{-1}$ is proportional to $\Omega^{-1}$ and thus invertible.}  as was anticipated in \cite{Gualtieri:2003dx}. As such, this generalizes the notion of A branes on K\"ahler manifolds. An important difference with respect to the purely chiral case, is that in this symplectic case, the boundary potential $W$ is essentially fixed by consistency requirements. This is in stark contrast with B branes, where $W$ is completely independent of $V$ and serves in a very straightforward way as a potential for the gauge field living on the brane. 

This leaves only the issue of adding chiral fields to the mix. In \cite{Sevrin:2008tp} this was solved in the case where only chiral and twisted chiral fields are present. As we hinted at before, a nice geometric interpretation like the one in the previous paragraph is still lacking for this case. Nevertheless, very concrete boundary conditions were obtained also in this case. For example, if Neumann conditions are chosen in all twisted chiral directions, one can combine the boundary constraints on the twisted chiral fields with the constraints on the chiral fields along the brane to obtain
\begin{eqnarray}
\hat D X^M = {\cal K}^M{}_N DX^N, \quad
\mbox{where} \quad
{\cal K} = \left( \begin{array}{cc}
            J & 0 \\
            L & K
            \end{array} \right) \label{calK},
\end{eqnarray}
and the upper (lower) components of $X^N$ are (twisted) chiral fields. Notice that nonzero $L$ now leads to components of $F$ with one leg in chiral and one leg in twisted chiral directions. It seems very likely that the same general structure persists once we include semi-chiral fields. 

\section{Outlook and applications}

The full set of possible boundary conditions for models including the three types of superfield is very close to being fully understood and will appear in a forthcoming paper \cite{wip}. A complete understanding of the geometry behind these boundary conditions, possibly with the help of generalized complex geometry, is certainly desirable and being investigated. Here, a more geometric understanding of the role of the boundary potential $W$ could be very important. Once a full classification is obtained, we have the necessary tools for studying more general -- e.g. six-dimensional -- examples as well as duality transformations between them. 

In a second part of this write-up, we focus on some four-dimensional examples and applications of this general formalism. In particular, we discuss  3-branes on the WZW model on $SU(2)\times U(1)$. This is then used as a starting point for using the power of superspace to perform explicit T-duality transformations which map these branes to various branes on K\"ahler manifolds. A particularly interesting application of this is the construction of new examples of coisotropic branes. We show how to for instance construct a maximally coisotropic brane on a K\"ahler manifold which not hyperk\"ahler. 

\begin{acknowledgement}
We thank Ulf Lindstr\"om, Martin Ro\v{c}ek and Maxim Zabzine for useful discussions and suggestions.
All authors are supported in part by the European Commission FP6 RTN
programme MRTN-CT-2004-005104. AS and WS
are supported in part by the Belgian Federal Science Policy Office through
the Interuniversity Attraction Pole P6/11,  and in part by the
``FWO-Vlaanderen'' through project G.0428.06. AW is supported in part by grant 070034022
from the Icelandic Research Fund.
\end{acknowledgement}

\end{document}